\documentclass[onecolumn,12pt]{article}
\usepackage{graphicx}
\begin{document}
\title{Extended GWS Model Using Left-Right Symmetry and Its Prediction on Neutrino Mass}
\maketitle
\begin{center}
\textbf{Asan Damanik}\footnote{E-mail:d.asan@lycos.com}\\
Department of Physics, Sanata Dharma University, \\ Kampus III USD Paingan, Maguwoharjo, Sleman, Yogyakarta, Indonesia\\
and Department of Physics, Gadjah Mada University,\\  Bulaksumur, Yogyakarta, Indonesia.\\
\end{center}
\begin{center}
\textbf{Mirza Sartiawan} and \textbf{Arief Hermanto}\\
Department of Physics, Gadjah Mada University,\\  Bulaksumur, Yogyakarta, Indonesia.\\
\end{center}
\begin{center}
\textbf{Pramudita Anggraita}\\
Center for Accelerator Technology and Material Process,\\ National Nuclear Energy Agency (BATAN), Yogyakarta, Indonesia.
\end{center}

\abstract{We evaluate the predictions of the left-right symmetry model based on $SU(2)\otimes U(1)$ gauge group with one bidoublet and one doublet Higgs fields on the parity violation, charged leptons and neutrinos masses.  Parity violation in the weak interaction is due to the very large $M_{W_{R}}$ mass compare to the $M_{W_{L}}$ mass.  The charged leptons acquire a Dirac mass via ordinary Higgs mechanism, meanwhile the neutrinos acquire a very small Dirac masses via seesaw-like mechanism with values $m_{1}=0.000043~\mbox{eV},\;
m_{2}=0.008888~\mbox{eV}$, and $m_{3}=0.014948~\mbox{eV}$ when $\rho=8.4\times 10^{-11}$, or $m_{1}=0.000014~\mbox{eV},\;m_{2}=0.003032~\mbox{eV}$, and $m_{3}=0.050990~\mbox{eV}$ when $\rho=2.9\times 10^{-11}$.

\maketitle

\section{Introduction}

One of the models which is very successful phenomenologically in explaining both electromagnetic and weak interaction in a unified theory is the Glashow-Weinberg-Salam (GWS) model.  Even though the GWS model was successful phenomenologically, but it still far from a complete theory because the GWS model blinds to many fundamental problems such as neutrino mass and fermions (lepton and quark) mass hierarchy.  Recent experimental data on the atmospheric and solar neutrinos indicate strongly that the neutrinos have masses and mixed up one another \cite{Fukuda98,Fukuda99, Fukuda01, Toshito01, Giacomelli01, Ahmad02, Ahn03}.  

Many models have been proposed to extend the GWS model.  One of the interesting models is the left-right symmetry model based on $SU(2)_{L}\otimes SU(2)_{R}\otimes U(1)$ gauge group which is proposed by Senjanovic and Mohapatra \cite{Senjanovic75}.  Using the Higgs multiplets $X_{L}\equiv (\frac{1}{2},0,1)$, $X_{R}\equiv (0,\frac{1}{2},1)$, and $\sigma\equiv (\frac{1}{2},\frac{1}{2},0)$,  Senjanovic and Mohapatra found that the Higgs potential will be minimum when we choose the asymmetric solution to the doublet Higgs fields vacuum expectation values.  Within this scheme, the presence of the spontaneous parity violation at low energy arises naturally, the dominant V-A interaction at low energy can be understood, and the neutrino has mass which could be arbitrarily small.

The spontaneous breakdown of parity in a class of gauge theories have also been investigated by Senjanovic \cite{Senjanovic79}.  The left-right model for the quark and lepton masses without a scalar bidoublet has been proposed which can predict a small neutrino mass via a double seesaw mechanism \cite{Brahmachari03}.  But, this left-right model without bidoublet Higgs leads to a non-renormalizable theory.  Any viable gauge model of electroweak interactions must give an answer to the two quite different problems: (i) the breaking of symmetry from the full gauge group into electromagnetic Abelian group $U(1)_{em}$ which give a mass to the gauge bosons and then explain the known structure of weak interactions, and (ii) the mass matrices for fermions \cite{Siringo04}.

Motivated by the rich contents of the left-right symmetry model, in this paper we evaluate the structure of electroweak interactions and  neutrinos masses by using the left-right symmetry model which is based on $SU(2)\otimes U(1)$ with one bidoublet and one doublet Higgs fields.  In this model, both left and right fermion fields are represented as doublets of $SU(2)$ group.  In Section 2 we introduce explicitly our model and take the minimum value of Higgs potential as the constraint in choosing the Higgs vacuum expectation values.  In section 3, we evaluate the gauge bosons masses and its implications to the structure of electroweak interaction which is known dominant V-A interaction. In section 4 we evaluate the leptons masses and calculate explicitly the neutrinos masses. In section 5, we calculate the absolute neutrinos masses as the implications of the model, and finally in section 6 we present a conclusion.

\section{The Model}

We use a left-right symmetry model based on $SU(2)\otimes U(1)$ gauge group.  Both left- and right-handed fermions fields are represented as doublets of $SU(2)$
\begin{eqnarray}
\psi_{L}=\bordermatrix{&\cr
&\nu_{l}\cr
&l^{-}}_{L},\: \psi_{R}=\bordermatrix{&\cr
&\nu_{l}\cr
&l^{-}}_{R},
 \label{4.3}
\end{eqnarray}
where $l=e,\mu,\tau$.

To break the symmetry, we introduce two Higgs fields (one bidoublet ($\Phi$) and one doublet ($\phi$) of $SU(2)$ gauge group) as follow
\begin{eqnarray}
\Phi=\bordermatrix{& &\cr
&w^{0} &x^{+}\cr
&y^{-} &z^{0}\cr},\  \phi=\bordermatrix{&\cr
&k^{0}\cr
&j^{+}\cr}.
 \label{4.4}
\end{eqnarray}
The bidoublet Higgs field $\Phi$ transform as $
\Phi\longrightarrow U\Phi U^{\dagger}$, and  $\phi\longrightarrow U\phi$, where $U$ is an unitary matrix.  We assume that the bidoublet Higgs field is blind to the helicity.  In this model, we put the $W_{\mu}^{a}$ ($a=1,2,3$) vector bosons in $SU(2)$ space which undergoes interactions with both Higgs and fermions fields.  We assume that the $W_{\mu}^{a}$ to be projected into two components $W_{1\mu}^{a}$ and $W_{2\mu}^{a}$ due to its interactions with Higgs fields $\Phi$ and $\phi$ respectively.  Meanwhile, the fermions sector know the left-right symmetry and then the $W_{\mu}^{a}$ to be projected to two components $W_{\mu L}^{a}$ and $W_{\mu R}^{a}$ due to its interaction with  $\psi_{L}$ and $\psi_{R}$ respectively.  The $W_{\mu R}^{a}$ and $W_{\mu L}^{a}$ are related to the $W_{2\mu}^{a}$ and $W_{1\mu}^{a}$ as follow 
\begin{eqnarray}
W_{\mu R}^{a}=W_{1\mu}^{a}\cos \zeta+W_{2\mu}^{a}\sin \zeta,\nonumber\\
W_{\mu L}^{a}=-W_{1\mu}^{a}\sin \zeta+W_{2\mu}^{a}\cos \zeta,
 \label{LRa}
\end{eqnarray}
where $\zeta$ is the mixing angle which is related to the parity violation.  Since at high energy there is no parity violation, we then have $\zeta=45^{o}$.

The Lagrangian density for kinetic energy and its self interactions $\mathcal{L_{B}}$ in this model is given by
\begin{eqnarray}
\mathcal{L}_{B}=-\frac{1}{4}(W_{\mu\nu}^{a}W^{a\mu\nu}+B_{\mu\nu}B^{\mu\nu}),
 \label{4.7a}
\end{eqnarray}
where
\begin{eqnarray}
W_{\mu\nu}^{a}=\partial_{\mu}W_{\nu}^{a}-\partial_{\nu}W_{\mu}^{a}-g\epsilon^{abc}W_{\mu}^{b}W_{\nu}^{c},\:
B_{\mu\nu}=\partial_{\mu}B_{\nu}-\partial_{\nu}B_{\mu}.
\end{eqnarray}
To simplify notation, we do not write the index $a$ further more.

The Lagrangian density for kinetic energy of leptons and quarks and its interactions with bosons ($\mathcal{L}_{k}$) read
\begin{eqnarray}
\mathcal{L}_{mB}=Tr\left|(i\partial_{\mu}-\frac{g}{2}\tau W_{1\mu})\Phi-\frac{g}{2}\Phi\tau W_{1\mu}-g'\frac{I}{2}B_{\mu}\Phi\right|^{2}\nonumber\\
+\left|(i\partial_{\mu}-\frac{g}{2}\tau W_{2\mu}-g'\frac{I}{2}B_{\mu})\phi\right|^{2}
-\mathcal{V}(\Phi,\phi),
 \label{4.7c}
\end{eqnarray}
where $I=B-L$, and the Lagrangian density for fermions masses and its couplings to Higgs fields ($\mathcal{L}_{mf}$) given by
\begin{eqnarray}
\mathcal{L}_{mf}=-m_{q_{\alpha\beta}}\bar{q}_{Li}^{\alpha}q_{Rj}^{\beta}+G_{q_{\alpha\beta}}\bar{q}_{Li}^{\alpha}\Phi_{ij}^{\alpha\beta}q_{Rj}^{\beta}-m_{l_{\alpha\beta}}\bar{\psi}_{Li}^{\alpha}\psi_{Rj}^{\beta}+G_{q_{\alpha\beta}}^{'}\bar{q}_{Li}^{\alpha}\tau_{2}\Phi_{ij}^{\alpha\beta*}\tau_{2}q_{Rj}^{\beta}\nonumber\\-m_{l_{\alpha\beta}}\bar{\psi}_{Li}^{\alpha}\psi_{Rj}^{\beta}+G_{l_{\alpha\beta}}\bar{\psi}_{Li}^{\alpha}\Phi_{ij}^{\alpha\beta}\psi_{Rj}^{\beta}+G_{l_{\alpha\beta}}^{'}\bar{\psi}_{Li}^{\alpha}\tau_{2}\Phi_{ij}^{\alpha\beta*}\tau_{2}\psi_{Rj}^{\beta}+h.c.
 \label{4.7d}
\end{eqnarray}
In this model, we assume that $G_{l_{\alpha\beta}}>>G_{l_{\alpha\beta}}^{'}$.  According to the requirement that $Q\left\langle\Phi\right\rangle=0, \:Q\left\langle\phi\right\rangle=0$, where $Q$ is the electric charge operator, in this extended model which the bidoublet and doublet Higgs fields transform as $(\textbf{2}\times\textbf{2}^{*},0)$ and $(\textbf{2},-1)$ under $SU(2)\otimes U(1)$ gauge group respectively.  The Lagrangian density for Higgs potential is given by
\begin{eqnarray}
\mathcal{V}(\Phi,\phi)=A(\phi^{+}\phi)+BTr(\Phi^{+}\Phi)+C(\phi^{+}\phi)^{2} \ \ \ \ \ \ \ \ \ \ \ \ \ \ \ \ \nonumber\\ +DTr(\Phi^{+}\Phi)^{2}+E(\phi^{+}\Phi^{+}\Phi\phi)+(F\phi^{+}\Phi\phi+h.c).
 \label{4.8}
\end{eqnarray}
where $A,\:B,\:C,\:D,\:E$, and $F$ are parameters.  Choosing the vacuum expectation values of Higgs fields in eq. (\ref{4.4}) to be
\begin{eqnarray}
\left\langle \Phi\right\rangle=\bordermatrix{& &\cr
&0 &0\cr
&0 &z\cr},\  \left\langle \phi\right\rangle=\bordermatrix{&\cr
&k\cr
&0\cr},
 \label{4.9}
\end{eqnarray}
then Higgs potential density read
\begin{eqnarray}
\mathcal{V}=Ak^{2}+Bz^{2}+Ck^{4}+Dz^{4}.
 \label{4.10}
\end{eqnarray}

The values of $z$ and $k$ which give the $\mathcal{V}$ to be minimum can be obtained from $\frac{\partial \mathcal{V}}{\partial z}=0$ and $\frac{\partial \mathcal{V}}{\partial k}=0$.  Choosing the expectation values of the Higgs fields as in eq. (\ref{4.9}),  the fifth and sixth terms in eq. (\ref{4.8}) do not contribute to Higgs potential density  $\mathcal{V}(z,k)$.  Thus, the minimum value of the $\mathcal{V}(z,k)$ depends only on the values of parameters $A,\;B,\;C$, and $D$.  If $A,\;B<0$, then potential density will be minimum  when the $z$ and $k$ values are given by
\begin{eqnarray}
z=\sqrt{\frac{B}{2D}},\;\;k=\sqrt{\frac{A}{2C}}.
 \label{zk}
\end{eqnarray}
From eq. (\ref{zk}) we can see that the values of $z$ and $k$ are independent from each other.

\section{The Gauge Bosons Masses}

From eq. (\ref{4.7c}), the mass term for gauge bosons read 
\begin{eqnarray}
\mathcal{L}_{mB}=Tr\left|-\frac{g}{2}\tau W_{1\mu}\Phi-\frac{g}{2}\Phi\tau W_{1 \mu}-g'\frac{I}{2}B_{\mu}\Phi\right|^{2}
+\left|(-\frac{g}{2}\tau W_{2\mu}-g'\frac{I}{2}B_{\mu})\phi\right|^{2}.
 \label{4.12}
\end{eqnarray}
As we know from the experimental facts that the parity is violated maximally at low energy, then we can put $\zeta =0$ to accommodate this parity violation.  Thus, at low energy, we then have
\begin{eqnarray}
W_{\mu R}^{a}=W_{1\mu}^{a},\;
W_{\mu L}^{a}=W_{2\mu}^{a}.
 \label{LRa1}
\end{eqnarray}

Inserting the value of $B-L=0$ for bidoublet and -1 for doublet with the vacuum expectation values in (\ref{4.9}), at low enegy, the gauge bosons mass term reads
\begin{eqnarray}
\mathcal{L}_{mB}=\frac{g^{2}z^{2}}{2}\left[\left(W_{1\mu}^{1}\right)^{2}+\left(W_{1\mu}^{2}\right)^{2}\right]+g^{2}z^{2}\left(W_{1\mu}^{3}\right)^{2}+\frac{g^{2}k^{2}}{4}\left[\left(W_{2\mu}^{1}\right)^{2}+\left(W_{2\mu}^{2}\right)^{2}\right]\nonumber\\+\frac{g^{2}k^{2}}{4}\left(W_{2\mu}^{3}\right)^{2}+\frac{g'^{2}k^{2}}{4}\left(B_{\mu}\right)^{2}-\frac{gg'k^{2}}{2}W_{2\mu}^{3}B_{\mu},
 \label{4.14}
\end{eqnarray}
where the index $a$ has been dropped for simplicity.
If we introduce a new physical quantities $W_{\mu R}^{\pm}$ and $W_{\mu L}^{\pm}$ as
\begin{eqnarray}
W_{\mu R}^{\pm}=\frac{1}{\sqrt{2}}\left(W_{1\mu}^{1}\mp iW_{1\mu}^{2}\right),\;
W_{\mu L}^{\pm}=\frac{1}{\sqrt{2}}\left(W_{2\mu}^{1}\mp iW_{2\mu}^{2}\right),
\end{eqnarray}
then eq. (\ref{4.14}) becomes
\begin{eqnarray}
\mathcal{L}_{mB}=g^{2}z^{2}W_{\mu R}^{+}W_{R}^{\mu-}+\frac{g^{2}k^{2}}{2}W_{\mu L}^{+}W_{L}^{\mu-}+g^{2}z^{2}\left(W_{1\mu}^{3}\right)^{2}\nonumber\\+\frac{g^{2}k^{2}}{4}\left(W_{2\mu}^{3}\right)^{2}+\frac{g'^{2}k^{2}}{4}\left(B_{\mu}\right)^{2}-\frac{gg'k^{2}}{2}W_{2\mu}^{3}B^{\mu}.
 \label{4.14a}
\end{eqnarray}
From eq. (\ref{4.14a}) one can see that the boson $W_{1\mu}^{3}$ is independent, but the $W_{2\mu}^{3}$ related to the $B_{\mu}$ boson.  The unrelated of the $W_{1\mu}^{3}$ boson to the $B_{\mu}$ boson due to the bidoublet Higgs fields quantum numbers $I=B-L=0$.  Therefore we put this $W_{1\mu}^{3}$ boson independently.

The mass terms for charged bosons can be obtained from eq. (\ref{4.14a}) in $W_{\mu L}^{\pm}$, and $W_{\mu R}^{\pm}$ basis.  That charged bosons masses read
\begin{eqnarray}
M_{W_{R}}=gz,\:
M_{W_{L}}=\frac{gk}{\sqrt{2}}.
 \label{mRmL}
\end{eqnarray}
From eq. (\ref{mRmL}), if $z<<k$, then we have $M_{W_{R}}>>M_{W_{L}}$.  
In our left-right symmetry model, both left- and right-charged current contribute to the effective Lagrangian ($\mathcal{L}_{eff}$) as 
\begin{eqnarray}
\mathcal{L}_{eff}^{LR}=-(\frac{4G_{F_{L}}}{\sqrt{2}}j_{L}^{\mu}j_{L\mu}^{\dagger}+\frac{4G_{F_{R}}}{\sqrt{2}}j_{R}^{\mu}j_{R\mu}^{\dagger}).
\end{eqnarray}
where $G_{F_{L/R}}$ is the Fermi coupling ($\frac{4G_{F_{L/R}}}{\sqrt{2}}=\frac{g^{2}}{2M_{W_{L/R}}^{2}}$).
Since $M_{W_{L}}<< M_{W_{R}}$, it implies that $G_{F_{L}}>> G_{F_{R}}$ which mean that at low energy level our model predicts the maximal parity violation.

The neutral gauge boson mass in eq. (\ref{4.14a}) are the third to sixth terms.  The third term can be separated from the other because this term is unrelated to the $B_{\mu}$ boson field.  This third term can be made to be independent as a new neutral massive boson that has mass $M_{W_{1}^{3}}=gz$.  The fourth, fifth, and the sixth terms in eq. (\ref{4.14a}) can be written in a matrix form as
\begin{eqnarray}
M^{2}=\bordermatrix{&W_{2\mu}^{3} &B_{\mu}\cr
W_{2\mu}^{3}&\frac{g^{2}k^{2}}{4}&-\frac{gg'k^{2}}{4}\cr
B_{\mu}^{3}&-\frac{gg'k^{2}}{4}&\frac{g'^{2}k^{2}}{4}\cr}.
 \label{4.16}
\end{eqnarray}

The mass matrix in eq. (\ref{4.16}) have zero determinant, thus we then have one eigenvalue whose value is zero and another one which has non-zero eigenvalue.   Explicitly, the transformation from ($W_{2\mu}^{3},B_{\mu}$) basis into ($Z_{\mu},A_{\mu}$) basis can be written as
\begin{eqnarray}
Z_{\mu}=\frac{gW_{2\mu}^{3}-g'B_{\mu}}{\sqrt{g^{2}+g'^{2}}}=W_{2\mu}^{3}\cos\theta_{W}-B_{\mu}\sin\theta_{W},\nonumber\\
A_{\mu}=\frac{g'W_{2\mu}^{3}+gB_{\mu}}{\sqrt{g^{2}+g'^{2}}}=W_{2\mu}^{3}\sin\theta_{W}+B_{\mu}\cos\theta_{W},
 \label{basis}
\end{eqnarray}
where $\theta_{W}$ is the Weinberg angle or the weak mixing angle.

The zero eigenvalue is associated to the $A_{\mu}$ gauge boson which is known as the electromagnetic field, and the non-zero eigenvalue is associated to the neutral weak gauge boson $Z_{\mu}$ with mass $M_{Z}$.  The square root of the eigenvalues of the mass matrix in eq. (\ref{4.16}) are the masses of the gauge boson $A_{\mu}$ and $Z_{\mu}$ as follow
\begin{eqnarray}
M_{A}=0,\;
M_{Z}=\frac{k}{2}\sqrt{g^{2}+g'^{2}}.
 \label{bosonnetral}
\end{eqnarray}

\section{The Leptons Masses}

If all particles acquire their masses via Higgs mechanism, the mass term of the form $m_{l}\bar{\psi}_{Li}^{\alpha}\psi_{Rj}^{\beta}$ should not contribute and the neutrino masses can be neglected due to the assumption that $G_{l_{\alpha\beta}}>>G_{l_{\alpha\beta}}^{'}$.  Thus, the charged lepton masses ($m_{e},\:m_{\mu},\:m_{\tau}$) arise only from the Lagrangian density in eq. (\ref{4.7d}), that is
\begin{eqnarray}
\mathcal{L}_{mf}=G_{l_{\alpha\beta}}\bar{\psi}_{Li}^{\alpha}\Phi_{ij}^{\alpha\beta}\psi_{Rj}^{\beta}\ \ \ \ \ \ \ \ \ \ \ \ \ \ \ \ \ \ \ \ \ \ \ \ \ \ \ \ \ \ \ \ \ \ \ \ \ \ \ \ \ \ \ \ \ \ \ \ \ \ \ \ \ \ \ \ \nonumber\\
=\bordermatrix{& & &\cr
&\bar{e}_{L} &\bar{\mu}_{L} &\bar{\tau}_{L}\cr}\bordermatrix{& & &\cr
&G_{l_{ee}}z^{ee} &G_{l_{e\mu}}z^{e\mu} &G_{l_{e\tau}}z^{e\tau}\cr
&G_{l_{\mu e}}z^{\mu e} &G_{l_{\mu\mu}}z^{\mu\mu} &G_{l_{\mu\tau}}z^{\mu\tau}\cr
&G_{l_{\tau e}}z^{\tau e} &G_{l_{\tau\mu}}z^{\tau\mu} &G_{l_{\tau\tau}}z^{\tau\tau}\cr}\bordermatrix{&\cr
&e_{R}\cr
&\mu_{R}\cr
&\tau_{R}\cr}.
 \label{4.21}
\end{eqnarray}

The experimental facts show that there is no mixing in the charged lepton sector.  It implies that the value of $G_{\alpha\beta}=0$ when $\alpha\neq\beta$, thus the charged lepton mass matrix read
\begin{eqnarray} 
m_{l}=\bordermatrix{&e_{L} &\mu_{L} &\tau_{L}\cr
e_{R}&G_{l_{ee}}z^{ee} &0 &0\cr
\mu_{R} &0 &G_{l_{\mu\mu}}z^{\mu\mu} &0\cr
\tau_{R} &0 &0 &G_{l_{\tau\tau}}z^{\tau\tau}\cr}
 \label{4.24}
\end{eqnarray}
where $z^{ee}=z^{\mu\mu}=z^{\tau\tau}=z$ is the vacuum expectation value of the $\Phi$.  The charged lepton mass matrix in eq. (\ref{4.24}) produces the charged leptons masses $m_{e}=G_{l_{ee}}z, m_{\mu}=G_{l_{\mu\mu}}z$, and $m_{\tau}=G_{l_{\tau\tau}}z$.

Due to the Higgs vacuum expectation values in eq. (\ref{4.9}), the neutrinos do not acquire a mass.  But, the mass term for the neutrinos can be obtained by considering the Lagrangian density which resulted from a seesaw-like interactions as shown in Fig. 1.  The seesaw-like interaction is possible due to the existence of the following term in Lagrangian density $F\phi^{\dagger}\Phi\phi$ and $G_{l_{\alpha\beta}}\bar{\psi}_{Li}^{\alpha}\Phi_{ij}^{\alpha\beta}\psi_{Rj}^{\beta}$.
\begin{figure}[h]
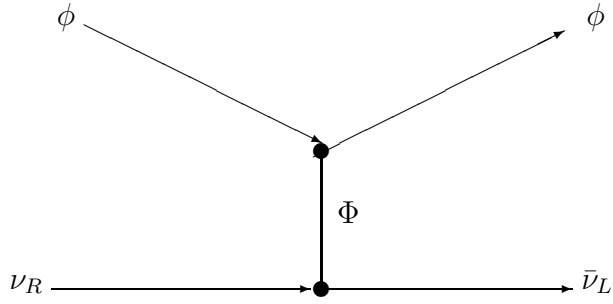

\begin{center}
$\nu_{R}\;$\vector(1,0){98}
\line(0,1){52}
\put(-100,100){$\phi\;$\vector(2,-1){90}\put(100,0){$\bar{\phi}$}}
\put(0,52){\circle*{6}}
\put(-3,50){\vector(2,1){95}}\put(6,25){$\Phi$}
\put(0,0){\circle*{6}}{\vector(1,0){95}$\;\bar{\nu}_{L}$}
\caption{Feynman diagram for a seesaw-like interaction}
\end{center}
\end{figure}

The Feynman diagram in Figure 1 gives an effective Lagrangian density ($\mathcal{L}_{ef}$)
\begin{eqnarray} \mathcal{L}_{ef}=\frac{F\phi^{\alpha}\bar{\phi}^{\beta}}{M_{\Phi}^{2}}G_{l_{\alpha\beta}}\bar{\nu_{L}^{\alpha}}\nu_{R}^{\beta},
 \label{ef}
\end{eqnarray}
where $F$ is the coupling between $\phi$ and $\Phi$ in eq. (\ref{4.8}), $G_{l_{\alpha\beta}}$ is the coupling of $\nu$ to $\Phi$, and $M_{\Phi}=\sqrt{B}$ is the bidoublet Higgs mass.  Since we have only one bidoublet and one doublet Higgs fields, then eq. (\ref{ef}) read
\begin{eqnarray} \mathcal{L}_{ef}=\frac{F\phi\bar{\phi}}{M_{\Phi}^{2}}G_{l_{\alpha\beta}}\bar{\nu_{L}^{\alpha}}\nu_{R}^{\beta},
 \label{Lef}
\end{eqnarray}
Inserting the vacuum expectation values of Higgs fields in eq. (\ref{4.9})) into (\ref{Lef}), we have
\begin{eqnarray} \mathcal{L}_{ef}=\frac{Fk^{2}}{M_{\Phi}^{2}}G_{l_{\alpha\beta}}\bar{\nu_{L}^{\alpha}}\nu_{R}^{\beta}.
 \label{Leff}
\end{eqnarray}

Since the values of $G_{\alpha\beta}=0$ when $\alpha\neq\beta$ as we have taken in the charged lepton sector, then eq. (\ref{Leff}) produces neutrino mass matrix in mass basis
\begin{eqnarray}
m=\frac{Fk^{2}}{M_{\Phi}^{2}}\bordermatrix{& & & \cr
&G_{l_{ee}}&0&0\cr
&0&G_{l_{\mu\mu}}&0\cr
&0&0&G_{l_{\tau\tau}}\cr}
 \label{man}
\end{eqnarray}

Since $M_{\Phi}^{2}=2Dz^{2}=B$, then eq. (\ref{man}) can be written to be
\begin{eqnarray}
m=\frac{Fk^{2}}{2Dz^{2}}\bordermatrix{& & & \cr
&G_{l_{ee}}&0&0\cr
&0&G_{l_{\mu\mu}}&0\cr
&0&0&G_{l_{\tau\tau}}\cr}=\bordermatrix{& & & \cr
&m_{1}&0&0\cr
&0&m_{2}&0\cr
&0&0&m_{3}\cr}.
 \label{man2}
\end{eqnarray}

Thus, neutrino mass arising from a seesaw-like mechanism, and the neutrino mass is a Dirac type with very small values.  From eq. (\ref{man2}) we have
\begin{eqnarray}
m_{1}=\frac{FG_{l_{ee}}k^{2}}{2Dz^{2}},\;\;
m_{2}=\frac{FG_{l_{\mu\mu}}k^{2}}{2Dz^{2}},\;\;
m_{3}=\frac{FG_{l_{\tau\tau}}k^{2}}{2Dz^{2}}.
 \label{mmn}
\end{eqnarray}
To obtain the normal hierarchy of neutrino masses ($m_{1}<m_{2}<m_{2}$) in mass basis, we must choose the values of the coupling $G_{l_{ee}}<G_{l_{\mu\mu}}<G_{l_{\tau\tau}}$ as we have taken in charged lepton sector.

\section{Absolute Neutrino Mass}

The charged lepton mass ($m_{l}$) is generated via ordinary Higgs mechanism, while the neutral lepton masses (neutrino masses) in this model is generated via a seesaw-like mechanism.  In the quark sector we have a mixing matrix $V_{CKM}$ which relates the quark in mass basis with the flavor basis, while in neutrino sector there is also a mixing matrix $V_{PMNS}$ which relates the neutrinos in mass basis ($\nu_{1},\nu_{2},\nu_{3}$) with neutrinos in the flavor basis ($\nu_{e},\nu_{\mu},\nu_{\tau}$).  The mixing matrix $V_{PMNS}$ can be used to obtain the neutrino mass matrix in flavor basis
\begin{eqnarray}
m_{\nu}=V_{PMNS}mV_{PMNS}^{T},
 \label{mnu}
\end{eqnarray}
where $m_{\nu}$ is the neutrino mass matrix in flavor basis and $m$ is the neutrino mass matrix in mass basis.  The explicit values of the neutrino mixing matrix obtained from the experiment is given by \cite{Gonzales05}
\begin{eqnarray}
\left|V_{PMNS}\right|=\bordermatrix{& & &\cr
&0.79-0.88 &00.47-0.61 &<0.20\cr
&0.19-0.52 &0.42-0.73 &0.58-0.82\cr
&0.20-0.53 &0.44-0.74 &0.56-0.81\cr}.
 \label{eq:6c}
\end{eqnarray}
According to the requirement that the $V_{PMNS}$ matrix must be orthogonal and the moduli values of its entries in simple numbers, then we put the $V_{PMNS}$ to be
\begin{eqnarray}
V_{PMNS}=\bordermatrix{& & &\cr
&-2/\sqrt{6} &1/\sqrt{3} &0\cr
&1/\sqrt{6} &1/\sqrt{3} &1/\sqrt{2}\cr
&1/\sqrt{6} &1/\sqrt{3} &-1/\sqrt{2}\cr}.
 \label{eq:6d}
\end{eqnarray}

If we use the $V_{PMNS}$ matrix in eq. (\ref{eq:6d}), then eq. (\ref{mnu}) becomes
\begin{eqnarray}
m_{\nu}=K\bordermatrix{& & & \cr
&a&b&b\cr
&b&c&d\cr
&b&d&c},
 \label{A}
\end{eqnarray}
where $K=\frac{FA}{12BC}$, $a=4G_{l_{ee}}+2G_{l_{\mu\mu}}$, $b=-2G_{l_{ee}}+2G_{l_{\mu\mu}}$, $c=G_{l_{ee}}+2G_{l_{\mu\mu}}+3G_{l_{\tau\tau}}$, and $d=G_{ee}+2G_{\mu\mu}-3G_{\tau\tau}$.  The neutrino mass matrix in flavor basis (eq. (\ref{A})) have the same pattern with the neutrino mass matrix proposed by Ma \cite{Ma02}.

From eqs. (\ref{4.24}) and (\ref{mmn}), we have
\begin{eqnarray}
m_{1}=\rho m_{e},\;\;m_{2}=\rho m_{\mu},\;\;m_{3}=\rho m_{\tau},
 \label{AD}
\end{eqnarray}
where $\rho =Fk^{2}/2Dz^{3}$.  From eq. (\ref{AD}) we can have
\begin{eqnarray}
\rho=\sqrt{\frac{m_{2}^{2}-m_{1}^{2}}{m_{\mu}^{2}-m_{e}^{2}}}
=\sqrt{\frac{\Delta m_{21}^{2}}{m_{\mu}^{2}-m_{e}^{2}}}.
 \label{rho1}
\end{eqnarray}
Global analysis on neutrino oscillation data gives \cite{Gonzales07}
\begin{eqnarray}
\Delta m_{21}^{2}=7.9_{-0.28}^{+0.27}\times 10^{-5}\ \mbox{eV}^{2},\;
\left|\Delta m_{31}^{2}\right|=2.6\pm 0.2\times 10^{-3}\; \mbox{eV}^{2},\nonumber \\
\theta_{12}=33.7\pm 1.3,\; \theta_{23}=43,3_{-3.8}^{+4.3},\ \theta_{13}=0_{-0.0}^{+5.2},\ \ \ \ \ \ \ \ \ \ \ \ 
 \label{eq:expne}
\end{eqnarray}
where $\Delta m_{21}^{2}=m_{2}^{2}-m_{1}^{2}$, $\left|\Delta m_{31}^{2}\right|=\left|m_{3}^{2}-m_{1}^{2}\right|$, $\theta_{ij}$ in degree.  The $\Delta m_{21}^{2}$ is obtained by analysing the neutrino oscillations data, while $\left|\Delta m_{31}^{2}\right|$ is obtained by analysing the atmospheric neutrinos oscillations data.
  If we use the values of $m_{e}=0,511$ MeV, and $m_{\mu}=105,66$ MeV together with the neutrino oscillation data in eq. (\ref{eq:expne}), we then have
\begin{eqnarray}
\rho=8.4\times 10^{-11}.
 \label{rho11}
\end{eqnarray}
Substituting the value of $\rho$ in eq. (\ref{rho11}) and $m_{e}=0,511$ MeV, $m_{\mu}=105,66$ MeV, and $m_{\tau}=1776,99$ MeV into eq. (\ref{AD}), we then have the neutrino masses
\begin{eqnarray}
m_{1}=0.000043~\mbox{eV},\;
m_{2}=0.008888~\mbox{eV},\;
m_{3}=0.014948~\mbox{eV}.
\end{eqnarray}

From eq. (\ref{AD}) it is also possible to have another value of $\rho$
\begin{eqnarray}
\rho=\sqrt{\frac{m_{3}^{2}-m_{1}^{2}}{m_{\tau}^{2}-m_{e}^{2}}}
=\sqrt{\frac{\Delta m_{31}^{2}}{m_{\tau}^{2}-m_{e}^{2}}}.
 \label{rho2}
\end{eqnarray}
Substituting the value of the $m_{e},\;m_{\tau}$ masses, and $\Delta m_{31}^{2}$ into eq. (\ref{rho2}), we then have
\begin{eqnarray}
\rho=2.9\times 10^{-11}
 \label{rho22}
\end{eqnarray}
If we use the value of $\rho$ in eq. (\ref{rho22}), $m_{e}=0,511$ MeV, $m_{\mu}=105,66$ MeV, and $m_{\tau}=1776,99$ MeV, we then have the neutrino masses as follow
\begin{eqnarray}
m_{1}=0.000014~\mbox{eV},\;
m_{2}=0.003032~\mbox{eV},\;
m_{3}=0.050990~\mbox{eV}.
\end{eqnarray}

\section{Conclusions}

Extension of the GWS model using left-right symmetry for fermions sector and based on $SU(2)\otimes U(1)$ gauge group with two Higgs fields of $SU(2)$ group (one bidoublet and one doublet) break the symmetry spontaneously from $ SU(2)\otimes U(1)$ down to $U(1)_{em}$.  The spontaneously symmetry breaking gives masses to four gauge bosons ($W_{L}^{\pm},\:W_{R}^{\pm},\:Z_{L},\:W_{3}$) with masses $M_{W_{L}},\;M_{W_{R}},\;M_{Z},\;M_{W_{3}}$ respectively, and one photon has zero mass.  The model proposed in this paper can also produces $M_{W_{L}}<<M_{W_{R}}$ which can be used to explain the parity violation.  The charged lepton mass can be generated via ordinary Higgs mechanism, while the neutrino mass is only generated via a seesaw-like mechanism.  The charged leptom masses have a normal hierarchy by taking the coupling values $G_{l_{ee}}<G_{l_{\mu\mu}}<G_{l_{\tau\tau}}$.  The charged lepton masses are related to the Higgs potential parameters.

In this model, a tiny neutrino masses with Dirac type can be generated via a seesaw-like mechanism which gives $m_{1}=0.000043~\mbox{eV},\;
m_{2}=0.008888~\mbox{eV}$, and $m_{3}=0.014948~\mbox{eV}$ when $\rho=8.4\times 10^{-11}$, or $m_{1}=0.00014~\mbox{eV},\;m_{2}=0.003032~\mbox{eV}$, and $m_{3}=0.050990~\mbox{eV}$ when $\rho=2.9\times 10^{-11}$.

\section*{Acknowledgments}

The first author (A.D.) would like to thank to the Graduate School of Gadjah Mada University Yogyakarta and the Dikti Depdiknas Indonesia for a BPPS Scholarship Program.


\begin{thebibliography}{99}
\bibitem{Fukuda98}
Y. Fukuda \textit{et al.}, \textit{Phys. Rev. Lett}. {\bf{81}}, 1158 (1998).
\bibitem{Fukuda99}
Y. Fukuda \textit{et al.},  \textit{Phys. Rev. Lett}. {\bf{82}}, 2430 (1999).
\bibitem{Fukuda01}
S. Fukuda \textit{et al.}, \textit{Phys. Rev. Lett.} {\bf 86}, 5651 (2001). 
\bibitem{Toshito01}
T. Toshito, \textit{et al.}, \textit{arXiv: hep-ex/0105023}.
\bibitem{Giacomelli01}
G. Giacomelli and M. Giorgini, \textit{arXiv: hep-ex/0110021}.
\bibitem{Ahmad02}
Q. R. Ahmad \textit{et al}., \textit{Phys. Rev. Lett}. {\bf89} (2002) 011301 [nucl-ex/0204008].
\bibitem{Ahn03}
M. H. Ahn \textit{et al.}, \textit{Phys. Rev. Lett}. {\bf{90}}(4), 041801 (2003).
\bibitem{Senjanovic75}
G. Senjanovic and R. N. Mohapatra, \textit{Phys. Rev}. {\bf{D12}}, 1502 (1975).
\bibitem{Senjanovic79}
G. Senjanovic, \textit{Nucl. Phys}. {\bf{B153}}, 334 (1979).
\bibitem{Brahmachari03}
B. Brahmachari, E. Ma, and U. Sarkar, \textit{Phys. Rev. Lett.}, {\bf{91}}, 011801 (2003).
\bibitem{Siringo04}
F. Siringo, \textit{Phys. Rev. Lett}. {\bf 92(11)},119101-1 (2004). 
\bibitem{Gonzales05}
M. C. Gonzales-Garcia, \textit{Phys. Scripta,} {\bf T 121} (2005).
\bibitem{Ma02}
E. Ma, \textit{Phys. Rev}. {\bf{D66}}, 117301(1-3) (2002).
\bibitem{Gonzales07}
M. C. Gonzales-Garcia and M. Maltoni, \textit{Phys. Rept.} {\bf 460} (2008).
\end{thebibliography}
\end{document}